# HEURISTIC BASED TASK SCHEDULING IN MULTIPROCESSOR SYSTEMS WITH GENETIC ALGORITHM BY CHOOSING THE ELIGIBLE PROCESSOR

Probir Roy[1], Md. Mejbah Ul Alam[1] and Nishita Das[2]

[1]Bangladesh University of Engineering and Technology, Bangladesh
`proy.cse@gmail.com, mejbah.alam@gmail.com`

[2]Chittagong University of Engineering and Technology, Bangladesh
`nishita_05@yahoo.com`

## ABSTRACT

*In multiprocessor systems, one of the main factors of systems' performance is task scheduling. The well the task be distributed among the processors the well be the performance. Again finding the optimal solution of scheduling the tasks into the processors is NP-complete, that is, it will take a lot of time to find the optimal solution. Many evolutionary algorithms (e.g. Genetic Algorithm, Simulated annealing) are used to reach the near optimal solution in linear time. In this paper we propose a heuristic for genetic algorithm based task scheduling in multiprocessor systems by choosing the eligible processor on educated guess. From comparison it is found that this new heuristic based GA takes less computation time to reach the suboptimal solution.*

## 1. INTRODUCTION

In multiprocessor based system the processing capability of processors may vary. The parallel tasks must be allocated into the processors such that the total completion time must be as less as possible. The optimal usage of processors is also expected. Again optimal task scheduling in multiprocessor systems is NP-complete [8], that is, finding optimal scheduling of tasks for multiprocessors is time consuming. These define the problem of task scheduling on multiprocessor systems to allocate a set of tasks to processors such that the optimal usage of processors and accepted computational time for scheduling algorithm are obtained. Genetic Algorithm, an evolutionary algorithm is used to find a suboptimal solution of the problem in considerable computation time. To reach the solution faster many heuristic based approach are used. By heuristic based approach the initial population is much closer to optimal solution. This results much less computation time in GA. In this paper, we propose a modification of heuristic approach of genetic algorithm method based on bottom-level by choosing the eligible processor for assigning the tasks which eventually decreases the computational time for finding the suboptimal schedule.

This paper is based on deterministic model, that is task dependencies and their execution time are known. The communication costs among the tasks are negligible and the numbers of processors are also fixed. The dependencies along with execution time of the tasks are represented by a Directed Acyclic Graph (DAG).

The remainder of this paper is organized as follows: brief summary of the related works, explanation of DAG in section 3, Heuristic based Genetic Algorithm explanation in section 4, the proposed improvement in section 5, the result of experimental studies are presented on section 6 and a conclusion and future work in section 7.

## 2. RELATED WORK

Since the beginning of the research on this field many approaches have been developed to solve the task scheduling on multiprocessor system. Some are heuristic based approach [9-11]; some rely on evolutionary approaches [12-14] and some follows the hybrid methods [15-18].

There are many heuristic based methods for solving multiprocessor task scheduling approach. The best heuristic approaches are based on task list technique [19]. In this technique, a list of descending priority ordered tasks is made. A task is selected from the head of the list and assigned to the processor. This method is also classified as static and dynamic. In static scheduling algorithms the list is not updated with new ordering at run time while the dynamic approaches do. Scheduling algorithms using t-level (top level) and b-level (bottom level) attributes for assigning priority to the processors have been proposed. There is another frequently used parameter ALAP (As Late As Possible) start time [20], which defines the longest possible execution time that a task can be postponed. List scheduling ISH (Insertion Scheduling) followed by DSH (Duplication Scheduling) that is a task duplication method has been also proposed [21]. There are several other Heuristic methods (Level-based Heuristics) [22] such as HLFET ((Highest Level First with Estimated Times), HLFNET (Highest Levels First with No Estimated Times), Random (the assigned tasks priority are random), SCFET (Smallest Co-levels First with Estimated Times) and SCFNET (Smallest Co-levels First with No Estimated Times), CP/MISF (critical path/most immediate successors first) [10], HNF (Heavy Node First) and WL(Weighted Length) [19]. All of these attributes act based on level concept in the DAG and without consideration of communication cost. Moreover, DF/IHS [10], EZ (Edge-zeroing) algorithm [23], LC (Linear Clustering) algorithm [24], DSC (Dominant Sequence Clustering) algorithm [25], MD (Mobility Directed) [27], DCP (Dynamic Critical Path) [11], ETF (Earliest Task First) [9] and greedy heuristics [26] are other heuristic methods.

These heuristic based methods are not considered as intense as before as they do not have good result in all cases. Therefore research on combinatorial optimization algorithms such as GA [2, 3,12, 13, 28], meta-heuristics and hybrid methods [5, 18] are going on.

## 3. THE DAG MODEL

A Directed Acyclic Graph (DAG) for tasks is the graph that represents the precedence constraints among the tasks along with their execution time. The DAG can be represented by a set G = {V,E} where V is the set of the task and E is the set of relations between the tasks. When the DAG is represented in graph, V represents the nodes and E represents the edges among the nodes. The computation cost of a task is represented the by the weight of the node and is denoted by W (Ti) where Ti is the ith task. Figure 1 represents a DAG of 11 tasks along with their precedence constraints. Each edge in the DAG represents the relationship between the tasks.[23]

If there is an edge Eij from task Ti to task Tj then task Ti precedes the task Tj that is Ti is the predecessor of Tj and Tj is successor of Ti. It is also represented by the Ti >= Tj . The height of a task height (Ti) can be represented by

$$height(Ti) = \begin{cases} 0 & if\ PRED(Ti) = \emptyset \\ & otherwise\ (1) \\ 1 + \max\ height(Tj) | Tj \in PRED(Ti) \end{cases}$$

where PRED (Ti) is the set of predecessors of the task Ti.

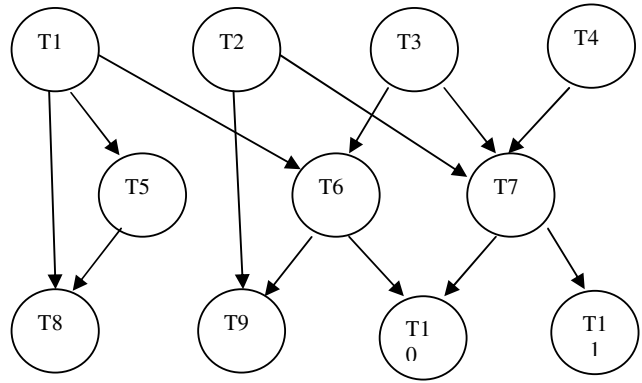

Figure 1. An example of a DAG

If a task Ti is assigned to a processor then ST (Ti) and FT (Ti) denote the start time and finishing time respectively. When all the tasks are scheduled, MAX {FT (Ti)} denotes the schedule length across all processors. The goal of scheduling is to minimize the MAX {FT (Ti)}.

In this paper we consider a set of processors P where all the processors are homogeneous i.e. all the processor will have same execution time to run a task individually. The number of processors is bounded and all the processors are connected with negligible communication cost.

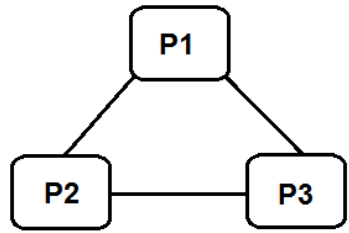

Figure 2 represents three fully connected homogeneous parallel systems.

Figure 2. Three fully connected machines

## 4. GENETIC ALGORITHM

A genetic algorithm is an evolutionary algorithm which generates near optimal solution of a problem by a guided random search method where elements (called individuals) in a given set of solutions (called population) are randomly combined and modified until some termination condition is achieved. The population evolves iteratively in order to improve a given cost function or fitness function of its individual [4]. In our case, the individuals are all the task-processor pairs and combinations of multiple individuals form the term population. The fitness of a population is MAX {FT (Pi)} for j =1, 2, 3, ..., Pn, where Pn is the number of processors and FT is the finishing time of the final task in processor Pi. The objective of GA is to find the minimum MAX {FT (Pi)}.

The initial step for a genetic algorithm is to get a set of initial population. Typically these populations are generated randomly. Then these populations are reviewed based on fitness function. Good individuals are replicated while bad individuals are removed. After selecting a pair of parent solution a crossover operation is performed to produce child solutions which preserve the characteristic of their parents. The mutation operation is randomly performed so that the random search algorithm does not stick in local minima. The mutation operation injects new characteristics in population to explore the uncovered area of the random search.

## 5. SCHEDULING ALGORITHMS

### 5.1. Heuristic Based Scheduling Algorithms

For the bounded number of processors (BNP) there are several scheduling algorithms i.e. HLFET algorithm, ISH algorithm, MCP algorithm, ETF algorithm, DLS algorithm. From them HLFET [22] is the simplest algorithms based on list scheduling technique. The basic idea of list scheduling is to make a scheduling list (a sequence of nodes for scheduling) by assigning them some priorities, and then repeatedly execute the following two steps until all the nodes in the graph are scheduled:

  (1) Remove the first node from the scheduling list;

  (2) Allocate the node to a processor which allows the earliest start-time.

There are various ways to determine the priorities of nodes. HLFET algorithm uses b-level based priority scheduling. The HLFET algorithm can be described as below

  (1) Calculate the static b-level (i.e., sl or static level) of each node.

  (2) Make a ready list in a descending order of static b-level. Initially, the ready list contains only the entry nodes. Ties are broken randomly.

Repeat

  (3) Schedule the first node in the ready list to a processor that allows the earliest execution, using the non-insertion approach.

  (4) Update the ready list by inserting the nodes that are now ready.

Until all nodes are scheduled.

### 5.2. Heuristic Based Genetic Algorithm

Performance of GA greatly depends on initial population as the more fit the initial population the faster it converges towards suboptimal solution.
Each individual of the initial population is generated through a minimum execution time or min-min heuristic along with b-level or t-level precedence resolution to avoid the problem of same execution time or completion time and same precedence. The problem of same execution time/completion time and precedence can occur in the homogeneous parallel system as all the processors take same execution time to execute one task.

The task to be scheduled for each iteration using b-level precedence resolution is determined by the following rules:[1,6]

(1) Sort the tasks according to their heights in ascending order.
(2) Sort the tasks with the same height according to their bottom-level in descending order.
(3) Repeat step 4 and step 5 until finish of all the tasks.
(4) Generate a permutation of processors.
(5) Assign tasks to processors in order.
(6) The above steps are repeated for the number of population size.

The length of all individuals in an initial population is equal to the number of tasks in the DAG. Following Table 1 represents the execution time and the priority of tasks' execution based on their bottom-level of the DAG presented in Figure 1.

| Task number | Execution time | Height | Bottom-level | Order of execution according to bottom-level |
|---|---|---|---|---|
| 1 | 50 | 0 | 72 | 1 |
| 2 | 1 | 0 | 41 | 4 |
| 3 | 10 | 0 | 50 | 3 |
| 4 | 20 | 0 | 60 | 2 |
| 5 | 20 | 1 | 21 | 7 |
| 6 | 2 | 1 | 22 | 6 |
| 7 | 20 | 1 | 40 | 5 |
| 8 | 1 | 2 | 1 | 11 |
| 9 | 20 | 2 | 20 | 9 |
| 10 | 19 | 2 | 19 | 10 |
| 11 | 20 | 2 | 20 | 8 |

Table 1: Execution time and priority of execution based on bottom-level

Figure 3 shows schedule for three processors based on the order of execution according to the bottom-level. From the figure we see that the total finish time based on the priority of the tasks' bottom-level is 92.

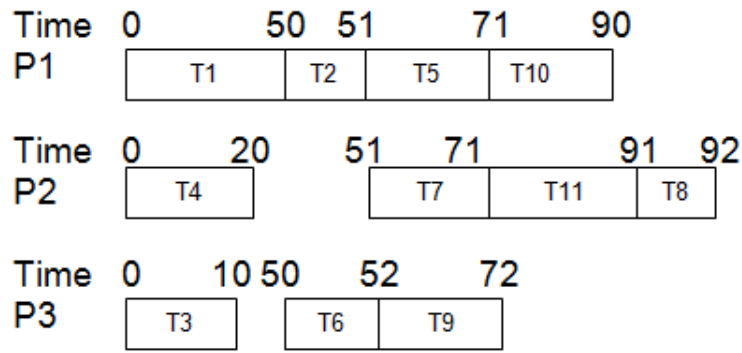

Figure 3: Schedule for three processors based on the Order of execution according to the bottom-level

## 5.3. Variation of HLFET heuristic over GA

In step 4 of Heuristic based GA using b-level precedence resolution a permutation of processors is performed. Instead of choosing processor by performing permutation of processors we propose the following algorithm.

Let the task mapping on the processors be represented by A {T, P} where T = {T1, T2, …, Tn}, a set of n tasks and P = {P1,P2,…,Pm}, a set of m processors. A{T, P} represents that task Ti $\epsilon$ T is mapped on processor Pk $\epsilon$ P. A set of unassigned tasks is represented as U = {T1, T2, …,Tn} where the tasks are sorted in descending order by b-level precedence. Tasks weight Tw is the execution time of task T. Processor's weight Pkw is the total execution time and waiting time of the processor so far.

$$Pk_w = \sum_{i=0}^{n} Ti_w + Wk$$

where Ti is assigned task on processor Pk and Tiw is the execution time of task Ti and Wk is the total waiting time of the processor Pk. PkTiw is the processors weight when the task Ti is assigned on processor Pk. Procedure LastTask(Pk) returns the last assigned task on processor Pk.

The pseudo code of our algorithm is:

```
Begin:
    While(U ≠ ϕ)
    Begin:
        Select first task Tj from unassigned task set U and remove the task from U.
        Find Pk from max(PkTiw, Ti ϵ PRED(Tj)).
        If(Pk ≠ ϕ && Ti = LastTask(Pk))
        Begin:
            Assign A{Tj, Pk}
            Pkw += Tjw
        End
        Else If(Pk ≠ ϕ)
        Begin:
            Find Pk = min (Pkw)
            Assign A{Tj, Pk}
            Pkw = max(PkTiw, Ti ϵ PRED(Tj)) + Tjw
        End
        Else
        Begin:
            Find Pk = min (Pkw)
            Assign A{Tj, Pk}
            Pkw += Tjw
        End
        End If
    End of While
End
```
Table 2: Variation of HLFET heuristic over GA

By applying this algorithm on the DAG of figure 3, we get the schedule for three processors system of figure 4.

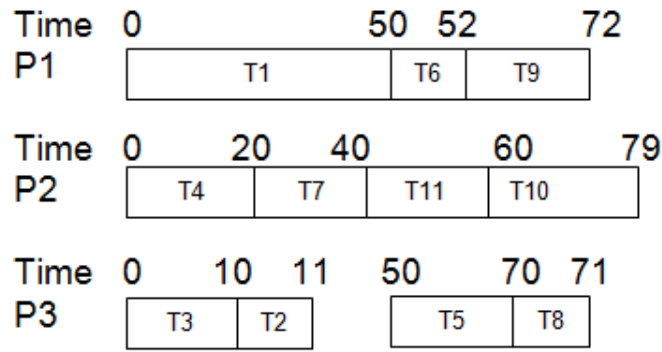

Figure 4: Schedule for three processors based on our proposed variation of HLFET heuristic over GA

From the figure we see that the total finish time of our algorithm is 79 where as the total finish time of round robin processor selection is 92.

## 6. EXPERIMENT RESULTS

Standard Task Graph (STG) has been used to benchmark the evaluation of multiprocessor scheduling algorithms [7]. All the task graphs in STG had been generated randomly without communication cost. As it is hard to evaluate on all the task graphs, we have chosen several task graphs randomly from the standard STG and performed the evaluation of the Elitism algorithm and our proposed scheduling algorithm. The tests have been run on 3, 4 and 16 processors with 50 and 100 tasks. For each set of test the test has been run for 10 times and from them average makespan - best makespan, average and best number of evaluations (to reach the termination condition) have been found.

Table 3 shows the simulation result of Elitism algorithm and our proposed algorithm in terms of average makespan, Best finish time, average and best number of evaluations to reach the termination condition.

|  | Elitism algorithm | | | | Proposed Algorithm | | | |
|---|---|---|---|---|---|---|---|---|
| Problem | Avg costs: | Best costs: | Avg evaluation | Best evaluation | Avg costs: | Best costs: | Avg evaluation | Best evaluation |
| 50tasks\Rand0100\4processors | 0.00084864 | 0.0008336 | 182.8 | 150 | 0.00085296 | 0.0008256 | 169.3 | 135 |
| 50tasks\Rand0100\16processors | 0.00154016 | 0.0013208 | 58.7 | 1 | 0.00149448 | 0.0011544 | 78.1 | 1 |
| 50tasks\Rand0069\4processors | 0.00212304 | 0.0020992 | 146.5 | 96 | 0.0019688 | 0.0019688 | 1 | 1 |
| 50tasks\Rand0069\16processors | 0.00226752 | 0.0022632 | 20.5 | 1 | 0.0019688 | 0.0019688 | 1 | 1 |
| 50tasks\Rand0019\4processors | 0.00071768 | 0.000704 | 158.9 | 131 | 0.00066264 | 0.000656 | 87.9 | 12 |
| 50tasks\Rand0019\16processors | 0.00088928 | 0.00078 | 190.3 | 165 | 0.00079704 | 0.0007704 | 158.3 | 81 |
| 50tasks\Rand0016\4processors | 0.00069064 | 0.0006736 | 151.9 | 108 | 0.00064984 | 0.0006392 | 58.4 | 1 |
| 50tasks\Rand0016\16processors | 0.00082856 | 0.0007992 | 173.8 | 128 | 0.00076744 | 0.000704 | 180.2 | 135 |
| 50tasks\Rand0002\4processors | 0.00050512 | 0.0004744 | 169 | 106 | 0.00049024 | 0.0004712 | 162.3 | 119 |
| 50tasks\Rand0002\3processors | 0.0000696 | 0.0000685 | 62.7 | 10 | 0.0000694 | 0.000069 | 46.7 | 5 |
| 50tasks\Rand0002\16processors | 0.0008328 | 0.0006888 | 190 | 171 | 0.00060336 | 0.000548 | 192.4 | 176 |
| 100tasks\Rand0100\4processors | 0.00121952 | 0.0011928 | 179.9 | 140 | 0.00122984 | 0.0011384 | 171.1 | 136 |
| 100tasks\Rand0100\16processors | 0.002604 | 0.002604 | 1 | 1 | 0.002604 | 0.002604 | 1 | 1 |
| 100tasks\Rand0069\4processors | 0.001816 | 0.0017928 | 147.1 | 68 | 0.00179592 | 0.001776 | 143.2 | 53 |
| 100tasks\Rand0069\16processors | 0.00355936 | 0.0033864 | 181.3 | 151 | 0.0031724 | 0.0029736 | 182.2 | 155 |
| 100tasks\Rand0019\4processors | 0.00205848 | 0.002024 | 147 | 64 | 0.00203688 | 0.002016 | 134 | 75 |
| 100tasks\Rand0019\16processors | 0.00429416 | 0.0039576 | 154.1 | 1 | 0.00386936 | 0.0034488 | 170.4 | 91 |
| 100tasks\Rand0016\4processors | 0.00190968 | 0.0018432 | 174.9 | 101 | 0.00185688 | 0.001784 | 159.4 | 122 |
| 100tasks\Rand0016\16processors | 0.00410888 | 0.0038808 | 164.8 | 1 | 0.00357968 | 0.0032568 | 171.6 | 97 |
| 100tasks\Rand0002\4processors | 0.00169768 | 0.0015488 | 171.9 | 139 | 0.00167768 | 0.0015448 | 169.3 | 140 |
| 100tasks\Rand0002\16processors | 0.002696 | 0.002696 | 1 | 1 | 0.002696 | 0.002696 | 1 | 1 |

Table 3: comparison between Elitism algorithm and our proposed algorithm

Figure 5 represents the comparison of average number of evaluations to reach the termination condition with Elitism Algorithm and our proposed algorithm. From the figure it is clear that our proposed algorithm takes much less time to reach the termination condition than Elitism Algorithm except the cases where there are 16 processors. While evaluating the figure 6 which represents the comparison of average make span we find that our algorithm performs much better than Elitism algorithm even for the 16 processors.

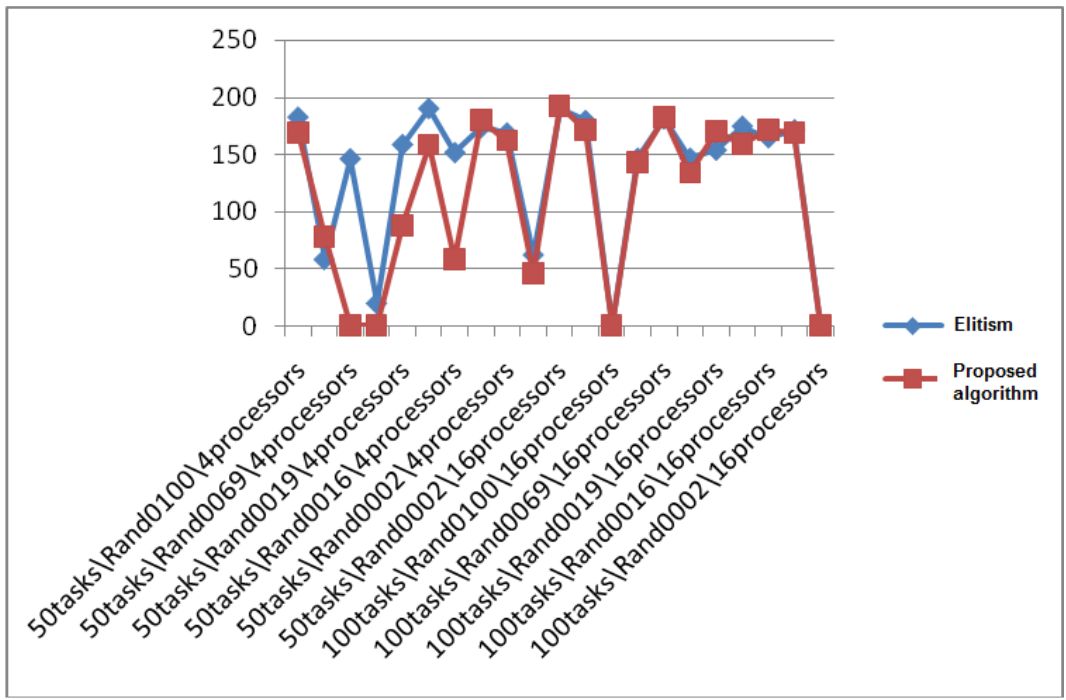

Figure 5: comparison of average number of evaluations to reach the termination condition with Elitism Algorithm and our proposed algorithm

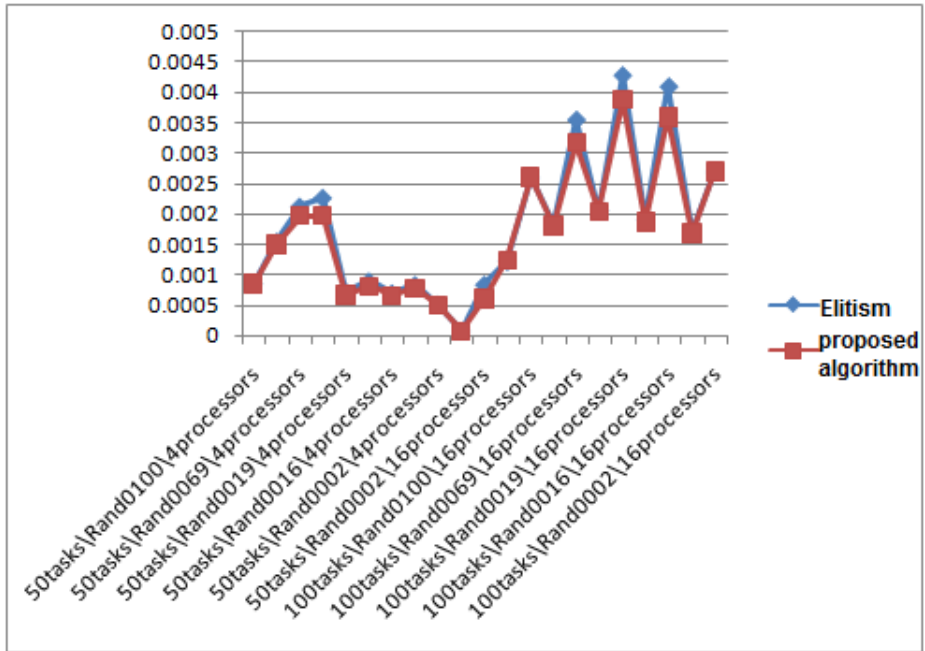

Figure 6: comparison of average makespan with Elitism Algorithm and our proposed algorithm

Figure 7 shows the steps of evaluation of a test case. In the test case there are 100 task with 16 processors. The test data is based on Rand069. The evaluation steps are represented along x-axis and the maximum makespan or cost is represented on y-axis. From the plot it is visible that the evaluation path of our algorithm is beneath of the Elitism algorithm. This means that our algorithm shedules the tasks with lower makespan quickly than the Elitism Algorithm.

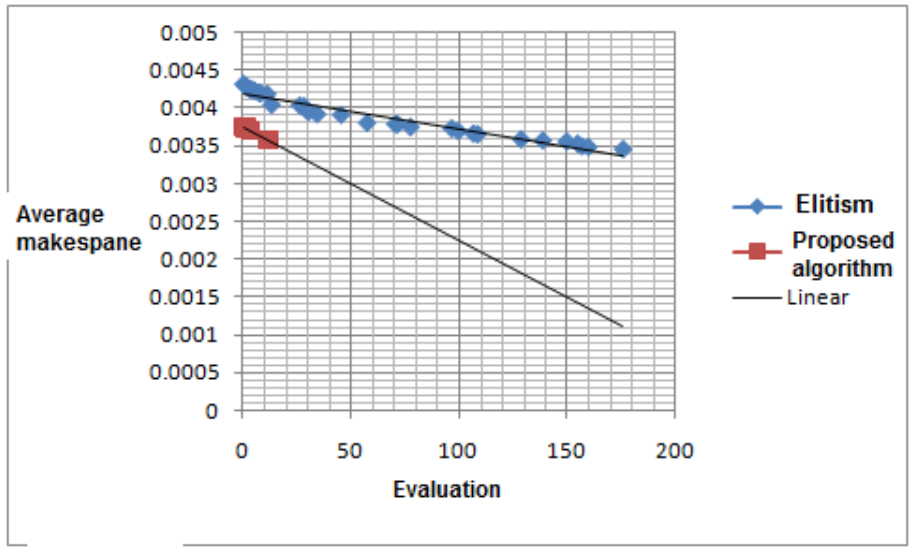

Figure 7: comparison of average makespan - evaluation with Elitism Algorithm and our proposed algorithm for 100 task with 16 processors (Rand069)

119

## 7. CONCLUSIONS

In this paper we have discussed the HLFET algorithm and Elitism Stepping method for task scheduling in multiprocessor systems with genetic algorithm. We also proposed a variation of HLFET algorithm and simulated the algorithm as well as Elitism stepping method with Standard Task Graph (STG). We have compared the simulation result and found that our proposed algorithm has better average makespan in smaller number of evaluations than Elitism Stepping method.

## REFERENCES

[1] Amir Masoud Rahmani, Mohammad Ali Vahedi, "A novel Task Scheduling in Multiprocessor Systems with Genetic Algorithm by using Elitism stepping method", Science and Research branch, Tehran, Iran, May 26, 2008.

[2] Auyeung, A., Gondra, I. and Dai, H.K. "Multiheuristic List Scheduling Genetic Algorithm for Task Scheduling", Proceedings of the Eighteenth Annual ACM Symposium on Applied Computing, ACM Press, pp. 721-724, 2003

[3] Yun Wen, Hua Xu, Jiadong Yang , (2010), " A heuristic-based hybrid genetic algorithm for heterogeneous multiprocessor scheduling ", Genetic And Evolutionary Computation Conference, pp. 729-736.

[4] Haupt, R.L., Haupt, S.E., Parallel genetic algorithms, John Wiley & Sons, 2004.

[5] Vahid Majid Nezhad1, Habib Motee Gader2 and Evgueni Efimov3, (2011),"A Novel Hybrid Algorithm for Task Graph Scheduling", IJCSI International Journal of Computer Science Issues, Vol. 8.

[6] Amir Masoud Rahmani and Mojtaba Rezvani, "A Novel Genetic Algorithm for Static Task Scheduling in Distributed Systems", International Journal of Computer Theory and Engineering, Vol. 1, No. 1, April 2009, 1793-8201..

[7] Standard task graph set is available online at: http://www.kasahara.elec.waseda.ac.jp/schedule

[8] M.R. Gary and D.S. Johnson, Computers and Imractability: A Guide to the Theorv of NP-Comnleteness. W.H. Freeman and Company, 1979.

[9] Hwang, J., Chow,Y.,Anger, A., Lee, C.: Scheduling precedence graphs in systemswith inter-processor communication times. SIAM J. Comput. 8(2), 244–257 (1989)

[10] Kasahara, H., Narita, S.: Practical multiprocessing scheduling algorithms for efficient parallel processing. IEEE Trans. Comput. 33, 1023–1029 (1984)

[11] Kwok, Y.-K., Ahmad, I.: Dynamic critical path scheduling: an effective technique for allocating task graphs to multiprocessors. IEEE Trans. Parallel Distrib. Syst. 7(5), 506–521 (1996)

[12] Hwang R.K., Gen M.: Multiprocessor scheduling using genetic algorithm with priority-based coding. In: Proceedings of IEEJ Conference on Electronics, Information and Systems, 2004

[13] Wu, A.S., Yu, H., Jin, S., Lin, K.-C., Schiavone, G.: An incremental genetic algorithm approach to multiprocessor scheduling. IEEE Trans. Parallel Distrib. Syst. 15(9), 824–834 (2004)

[14] Lee, Y.H., Chen, C.: A modified genetic algorithm for task scheduling in multi Processor systems. In: Proceedings of the NinethWorkshop on Compiler Techniques for High Performance Computing, 2003

[15] Sivanandam, S.N., Visalakshi, P., Bhuvaneswari, A.: Multiprocessor scheduling using hybrid particle swarm optimization with dynamically varying inertia. Int. J. Comput. Sci. Appl. 4(3), 95–106 (2007)

[16] Chen, H., Cheng, A.K.: Applying ant colony optimization to the partitioned scheduling problem for heterogeneous multiprocessors. Special Issue IEEE RTAS 2005Work-in-Progress 2(2), 11–14 (2005)

[17]     Ercan,M.F.: A hybrid particle swarm optimization approach for scheduling flow-shops with multiprocessor tasks. In: Proceedings of the International Conference on Information Science and Security, pp. 13–16 (2008)

[18]     M.Ebrahimi Moghaddam, M.R. Bnyadi , An Immune-based Genetic Algorithm with Reduced Search Space Coding for Multiprocessor Task Scheduling Problem,   International Journal of Parallel Programming (IJPP) , Springer, DOI 10.1007/s10766-011-0179-0, 2011

[19]     Shirazi, B., Wang, M., Pathak, G.: Analysis and evaluation of heuristic methods for static task scheduling. J. Parallel Distrib. Comput. 10(3), 222–232 (1990)

[20]     Kwok, Y.-K., Ahmad, I.: Dynamic critical path scheduling: an effective technique for allocating task graphs to multiprocessors. IEEE Trans. Parallel Distrib. Syst. 7(5), 506–521 (1996)

[21]     Kruatrachue, B., Lewis, T.G.: Duplication Scheduling Heuristic, a New Precedence Task Scheduler for Parallel Systems. Technical Report, Oregon State University (1987)

[22]     T.L. Adam, K. Clhandy and J. Dickson, "A Comparison of List Scheduling for Parallel Processing Systems," Communications of the ACM, vol. 17, no. 12, pp. 685-690, Dec. 1974.

[23]     Sarkar, V.: Partitioning and Scheduling Parallel Programs for Multiprocessors. MIT Press, Cambridge (1989)

[24]     Kim, S.J., Browne, J.C.: A general approach to mapping of parallel computation upon multiprocessor architectures. In: Proceedings of International Conference on Parallel Processing, pp. 1–8 (1988)

[25]     Yang, T., Gerasoulis, A.: List scheduling with and without communication delays. Parallel Comput. 19(12), 1321–1344 (1993)

[26]     Kermia, O., Sorel, Y.: A Rapid Heuristic for Scheduling Non-Preemptive Dependent Periodic Tasks onto Multiprocessor. ISCA PDCS, pp.1–6 (2007)

[27]     Wu, M.Y., Gajski, D.D.: Hypertool: a programming aid for message-passing systems. IEEE Trans. Parallel Distribut. Syst. 1(3), 330–343 (1990)

[28]     Hou, E.S.H., Ansari, N., Hong, R.: A genetic algorithm for multiprocessor scheduling. IEEE Trans. Parallel Distrib. Syst. 5(2), 113–120 (1994)